\documentclass[a4paper]{article}
\usepackage{graphicx}
\usepackage{amsmath}
\usepackage{hyperref}
\usepackage{color}

\begin{document}

\begin{center}

{\bf \Large  Altruism and reputation: cooperation within groups}\\[5mm]

{\large  P. Gawro\'nski, M. J. Krawczyk and K. Ku{\l}akowski$^*$}\\[3mm]

{\em

Faculty of Physics and Applied Computer Science,

AGH University of Science and Technology,

al. Mickiewicza 30, PL-30059 Krak\'ow, Poland

}

\bigskip

$^*${\tt kulakowski@novell.ftj.agh.edu.pl}

\bigskip

\today

\end{center}

\begin{abstract}
In our recent model, the cooperation emerges as a positive feedback between a not-too-bad reputation and an altruistic attitude.
Here we introduce a bias of altruism as to favorize members of the same group. The matrix $F_{i,j}$ of frequency of 
cooperation between agents $i$ and $j$ reveals the structure of communities. The Newman algorithm reproduces the initial bias. 
The method based on differential equations detects two groups of agents cooperating within their groups, leaving the uncooperative 
ones aside.

\end{abstract}

\noindent


\noindent


\section{Introduction}

Despite the widespread using of game theory \cite{neumor,straf} in social sciences, its range of applications 
remains disputable. The question, to what extent people are rational in their decisions, created a rich set of
 research points of view, including the concept of evolutionary thinking \cite{axel}. Axelrod discussed this
question in terms of the Prisoner's Dilemma (PD), a famous thought experiment which demostrates that according to 
 game theory in many situations a cooperation is not rational. Then, PD become a common method of real and computational 
experiments. To complete the list of references is beyond our chance and aim; we refer to \cite{ax6,szafa,brandt,fefi}.  
 Although the formulation of PD relies on the concepts of payoffs and rational choice, some recent experiments exceed 
this frame. In 2006 De Cremer and Stouten showed \cite{crem} that the conditions of cooperation are trust and some 
 common goal to achieve; the latter was measured with the 'Inclusion of Other with the Self' scale \cite{aron}. In the 
same year the results of Mulder et al. \cite{leti} indicated that sanctioning defection itself can weaken 
 the motivation of players to cooperate. These results, obtained within the scheme of the public good experiment 
\cite{ledy,urs}, show that a reliable theory of cooperation cannot be limited to the assumption of an idividual 
 and selfish rationality, but it should include some collectivistic attitudes, as for example preserving of social norms. 
The same conclusion can be drawn from the results of the ultimatum game, when played by members of different societies 
 \cite{hen}. \\

Here we are interested in a contribution of cooperation to the process of the formation of social groups. Although in 
a society this process usually runs in a longer timescale than a standard experiment, some information could be drawn 
 from a search of cooperation in already established groups. Here however, the outcome varies strongly from one particular 
case to another. Sometimes boundaries between groups seem to disappear, as those reported in \cite{mayama} between 
 Japanese and American students in Hokkaido. The data collected in Bosnia and North Causasus indicate, that the likelihood of 
intergroup trust and cooperation depends more on cultural and economic status than on personal - often crude - experience 
 \cite{cau}. In other cases, as the one with Australian and Singaporean students \cite{lohgal}, the results are less 
conclusive. The difference in timescale, noted above, was to some extent evaded in the PD experiment performed by Goette 
 et al. \cite{goe}. The groups investigated there were platoons of males formed for four-week period of officer training 
in the Swiss army. As individuals were randomly assigned to different platoons, the experiment was free from the 
confounding effect of self-selection into groups. The intra-group cooperation was found to be clearly stronger than the 
inter-group one. Also, individuals believed that members of their own platoons were more willing to cooperate. These 
 results are closely akin to those of \cite{crem}.\\

Recently we proposed a new model of cooperation, formulated without use of the concepts of payoff and utility \cite{kp}. 
In this model, the only model variables were reputations and altruisms of individual players. Both these variables entered
 to the expression of the probability of cooperation of one player with another. The only difference between the variables
was their time dependence; while the reputation varied in each game, the level of altruism remained constant. The aim of 
 this work is to apply the same model to the intra- and inter-group cooperation. Accordingly, we are going to admit that 
the reputation and the altruism can be different within the group and between the groups. \\

 The paper is organized as follows. Next section is devoted to the model, as formulated in \cite{kp}, and to a brief 
conclusion of its results for one group. In the same Section we explain how the model is generalized to the case of two 
groups. In Section 3 we report the way how the group structure is investigated; here we base on methods of \cite{negi,gos}. 
Numerical results are described in Section 4 and discussed in Section 5.

\section{The model}

The game is performed between $N$ agents placed at nodes of a fully connected graph. To each agent, two parameters are assigned:
reputation $W_i$ and altruism $\epsilon_i$, where $i=1,...,N$. The altruisms $\epsilon_i$ are random numbers from the homogeneous
distribution on the range $(-0.5,0.5)$, and these values remain constant during the game. Initial values of the reputations $W_i=0.5$.
Once an agent cooperates, hers/his reputation is transformed as $W_i \to (1+W_i)/2$; otherwise $W_i \to W_i/2$. At each time step 
a pair of agents $(i,j)$ is selected randomly from the whole set. The probability $P(i,j)$ that $i$ cooperates with $j$ is calculated 
as the fractional part of $W_j+\epsilon_i$; once this outcome is negative, $P(i,j)$ is set to zero, once it is larger than 1, $P(i,j)$
is set to 1 \cite{kp}.\\

\begin{figure}[ht] 
\centering
{\centering \resizebox*{12cm}{9cm}{\rotatebox{-90}{\includegraphics{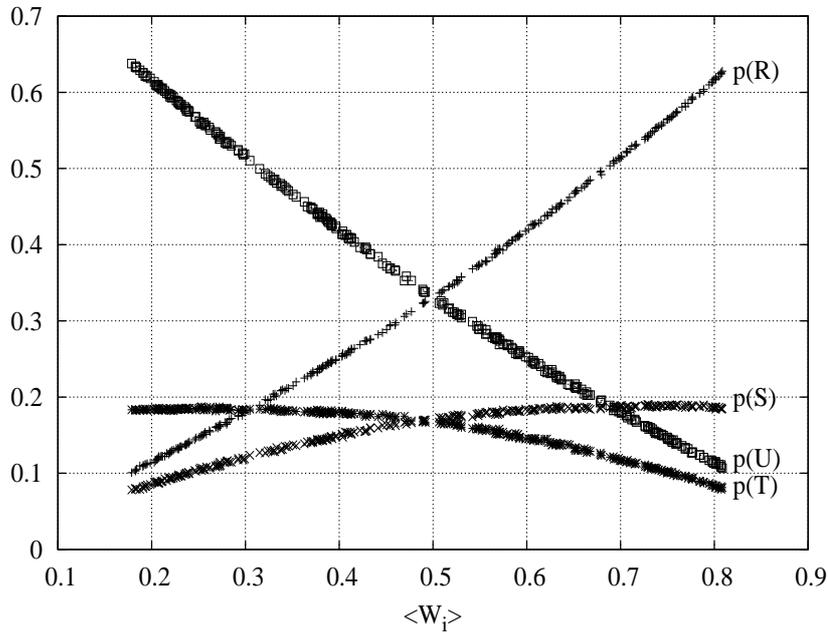}}}} 
\caption{
The probability of four different outcomes of PD against the time average of the reputation $<W_i>$.
}
\label{fig-2}
\end{figure}

The most important result of the model is presented in Fig. 1. What is shown there are frequencies of four situations met by an agent $i$ 
against the time average of the reputation $<W_i>$.  The situations are: 1. both $i$ and his partner cooperate, with the outcome R;
2. both $i$ and his partner defects, with the outcome U; 3. $i$ cooperates but his partner defects, with the outcome S; 4. $i$ defects but 
his partner cooperates, with the outcome T. As we see in Fig. 1, for agents with 
positive mean reputation $W$ the mutual cooperation is most frequent. On the contrary, agents with negative mean reputation $<W>$
defect in most cases: same do their partners. A similar plot was shown in \cite{kp}, with the difference that there, the altruism 
$\epsilon$ was used in the horizontal axis. We note that if we use the a temporary value of reputation $W_i$ and not its time average
$<W_i>$, the plot is so noisy that it is hard to see anything \cite{kp}.

\section{Group structure}

To investigate the group formation, here we divide the whole set of $N$ nodes into two halves, each of $N/2$ nodes and we introduce 
an additional bias parameter $\kappa$. The role of $\kappa$ is to enhance the intra-group cooperation and to weaken the inter-group one. 
With this purpose in mind, we apply the bias in two alternative ways: $\it i)$ by adding $\kappa$ to the initial values of reputations
of $i$ and $j$ if these nodes belong to the same group, and by reducing the initial values of reputations if $i$ and $j$ belong to different 
groups. For each $i$, this change of reputation $W_i$ is done only once, when $i$ plays for the first time. $\it ii)$ by a similar increase
(reduction) of attitudes $\epsilon_i,\epsilon_j$ by $\kappa$ when $i,j$ are in the same (different) groups. The parameter $\kappa$ varies from 
zero to 0.3.

The object to investigate now is the matrix $F_{i,j}$ which measures the frequency of cooperation of each two agents $i,j$. This matrix 
contains an information of a possible cluster structure of our network. Here we investigate it by two methods: the Newman algorithm (NM)		
\cite{negi} and our recent method by means of differential equations (DF) \cite{gos}. As all other methods designed for an identification 
the cluster structure \cite{sant}, these two methods have to extract this structure from an initial noised information. \\

The Newman algorithm is based on the time evolution of modularity $Q$  - a quantity which measures the departure of the actual 
graph structure from a random one. The formula is \cite{newm}

\begin{equation}
Q=\frac{1}{2m}\sum_{i,j}\Big[F_{i,j}-\frac{k_ik_j}{2m}\Big]\delta(c_i,c_j)
\end{equation}
where $k_i=\sum_jF_{i,j}$, $2m=\sum_{ij}F_{i,j}$ and $\delta(c_i,c_j)=1$ if $i,j$ belong to the same cluster; otherwise $\delta(c_i,c_j)=0$. 
We start from all $N$ nodes separated and we add links as to get maximal possible value of $Q$ at each step.  We accept the obtained cluster 
structure at the moment when $Q$ is maximal. For a more detailed description we refer to \cite{negi}.\\

Our algorithm with differential equations relies also on the modularity $Q$, but the time dependence of the connectivity matrix $C{i,j}$
is given by 

\begin{equation}
\frac{dC_{i,j}}{dt}=\Theta(C_{i,j})\Theta(1-C_{i,j})\sum_{k\ne i,j}(C_{i,k}C_{k,j}-\beta)
\end{equation}
where $\Theta(x)=0$ for $x<0$ and 1 in the opposite, and $\beta$ is a parameter. Our former numerical results suggest that the formalism works well if $\beta >0.4$. Here, the frequency matrix elements $F_{i,j}$ serve as initial values of $C_{i,j}$. Again, we break the time evolution at the moment when $Q$ gets its maximal value. The detailed description is given in \cite{gos}.

\section{Results}

Calculations are performed for $N$=300 nodes, i.e. for the frequency matrix $300\times 300$. The correlation between $\epsilon_i$ and $W_i$, 
initially equal to zero, increases and reaches its stationary value about 0.06 after $10^3$ games. All the results are then gathered after this
transient time. As a rule, $15\times 10^5$ games are played. The frequency matrix was obtained for both methods of using $\kappa$ to modify the initial reputation $W_i$ or the altruism $\epsilon_i$. For each of these methods and for the system of $N=300$ nodes we got four frequency matrices $F_{i,j}$ for $\kappa$= 0.0, 0.1, 0.2 and 0.3. These matrices can be used to produce the plots like our Fig. 1. There are only slight differences between these plots 
made for $\kappa$=0.0 and 0.3.\\

For both ways of using the bias $\kappa$, i.e. with modified reputation or altruism, NM gives the same results. This method gives always two 
clusters. For $\kappa=0$, their sizes are: 162 nodes, with mean $\epsilon$= -0.113,
and 138 nodes with mean $\epsilon$= 0.126. This could mean that the partition is roughly into defecting and cooperating agents. However, for 
$\kappa$=0.1, 0.2 and 0.3 the Newman algorithm reproduces exactly the division of the network into two groups  which were meant to cooperate 
by using the probabilities $P(i,j)$ with the preferences done by $\kappa$. In these groups, mean values of $\epsilon$ are close to zero.\\

\begin{figure}[ht] 
\centering
{\centering \resizebox*{12cm}{9cm}{\rotatebox{00}{\includegraphics{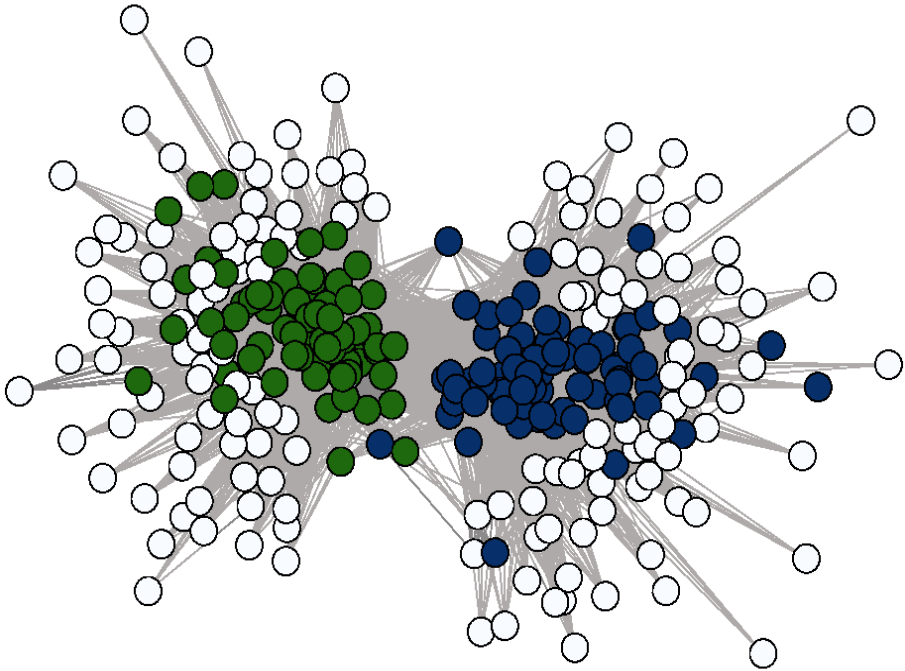}}}} 
\caption{
The cluster structure of the network of 300 nodes, described by the frequency matrix for the bias $\kappa=0.3$ (color online). For clarity of the picture, 
links where $F_{i,j}<0.5$ are omitted. Two clusters at the centre (grey and black, or green and blue online) are due to two internally cooperating groups, obtained with DF. As we see, there is also some cooperation between groups. The mean reputation of other nodes around the centre (white colour) is negative. The figure reveals that actually they cooperate to some extent, but almost exclusively within the initial division.
}
\label{fig-2}
\end{figure}

The results of DF \cite{gos} are different. In the case when the initial reputations $W_i$ are modified, we have one cluster: there is no partition.
When the altruisms $\epsilon_i$ are modified, for $\kappa$= 0.0, 0.1 and 0.2 and the parameter $\beta >0.1$ again there is no partition. For 
$\kappa$= 0.3 and all used values of $\beta$ (from 0.1 to 1.0) we get two clusters, with their sizes varying between 69 and 78 nodes; however, 
the size difference was not larger than 4 nodes. The obtained structure is visualised in Fig. 2. The altruism parameter $\epsilon$ 
averaged over each of those clusters, varied between 0.25 and 0.29. As we see, this partition captures cooperating agents in two groups. We 
deduce it from the mean values of $\epsilon$ for the clusters and from if their size. In the thermodynamic limit, we could expect the size to be 
$N/4$ and the mean altruism parameters to be 0.25. These limit values are close to those obtained in our simulation. \\

\section{Discussion}

Our question was, if the proposed model can account for the preference of cooperation within groups. The criterion designed here is, if the 
cooperation pattern, encoded in the frequency matrix $F$, is detectable as the structure of communities in the network. Both for the Newman 
algorithm \cite{negi} and for our method \cite{gos} the answer is affirmative. Although the answer is different in these two methods, in both 
cases we get a piece of nontrivial information. For $\kappa >0$, the Newman algorithm just reproduces the initial bias: the agents which should 
cooperate more willingly with each other do belong to the same cluster. This structure does not reflect the fact that many agents do not 
cooperate at all. For $\kappa$=0, the structure obtained with the Newman method is different: there are two clusters, one of them more 
cooperative than the other. However, this division is not strict; in our model a half of agents do cooperate. Basically, a most straightforward 
result for $\kappa=0$ would be that there is no division at all. 

The second method based on differential equations gives exactly this result - one cluster - for $\kappa=0$. However, the same result is obtained
also for $\kappa$ = 0.1 and 0.2. It is only for $\kappa$ = 0.3 when the algorithm detects two groups of agents cooperating 
within their groups, leaving the uncooperative agents aside. This result is the most expected one, as it gives an information both on the group structure
and on the cooperative attitude of members of these groups. The drawback of this method is that it gives the proper results only for large values
of the bias $\kappa$. As the parameter $\kappa$ measures the bias, the cluster structure obtained for $\kappa=0.3$ can be expected also for 
all larger $\kappa$. 

Summarizing, in our recent model \cite{kp} the coupling of altruism and reputation was described as a possible mechanism of successful cooperation.
Here we have shown that the same model can be generalized by adding the own-group preference to individual values of the parameter of altruism.
With this modification, the model is able to reproduce the process of group formation. The groups formed in this way can be detected by current methods
of search of cluster structure. In particular, the method based on differential equations \cite{gos} allowed to separate out two groups of agents cooperating within their groups. According to this method, uncooperative agents remain outside clusters.

\bigskip

{\bf Acknowledgements.} The research is partially supported within the FP7 project SOCIONICAL, No. 231288.

\end{document}